\documentclass[10pt,twocolumn,a4paper,conference]{IEEEtran}

\usepackage{amsmath}
\usepackage{amssymb}
\usepackage{amsfonts}
\usepackage[ansinew]{inputenc}

\usepackage{multirow}
\usepackage[nolist]{acronym}
\usepackage{acronym}
\usepackage[capitalize]{cleveref}
\usepackage{graphicx}
\usepackage{epstopdf}
\usepackage{microtype}
\usepackage{color}
\usepackage{subcaption}
\usepackage[font=footnotesize]{subcaption}
\usepackage[font=footnotesize]{caption}

\newenvironment{frcseries}{\fontfamily{frc}\selectfont}{}

\begin{acronym}
\acro{BS}{base station}
\acro{UE}{user}
\acro{MC-IC}{multicast interference channel}
\acro{IC}{interference channel}
\acro{MISO}{multiple-input single-output}
\acro{MIMO}{multiple-input multiple-output}
\acro{MBMS}{multimedia broadcast/multicast service}
\acro{eMBMS}{enhanced MBMS}
\acro{SDR}{semidefinite relaxation}
\acro{SNR}{signal to noise ratio}
\acro{SINR}{signal to interference and noise ratio}
\acro{PHY}{physical layer}
\acro{LBM}{leakage-based multicast beamforming}
\acro{RLBM}{relaxed LBM}
\acro{CSI}{channel state information}
\acro{iid}{independent and identically distributed}
\end{acronym}

\begin{document}

\title{Latency and Resource Utilization Analysis for V2X Communication over LTE MBSFN Transmission}
\author{Illia Safiulin$^\dagger$, Stefan Schwarz$^\dagger$, Tal Philosof$^\ddagger$ and Markus Rupp$^\dagger$ \\
$^\dagger$ Institute of Telecommunications, Technische Universität Wien \\
$^\ddagger$ Wireless Enablers Lab, General-Motors R\&D Advanced Technical Center Israel\\
Email: \{isafiuli,sschwarz,mrupp\}@nt.tuwien.ac.at, tal.philosof@gm.com
\vspace*{10pt}
}
\maketitle

\vspace*{-10pt}
\begin{abstract}  
In this paper, we investigate the performance of LTE Multicast-Broadcast Single-Frequency Networks (MBSFN). LTE-MBSFN is viewed as one of the most promising candidates for vehicular communications which can enhance reliability of vehicular application traffic. This is achieved  due to the possibility to efficiently support message exchange in-between vehicles by multicasting information to several vehicles in parallel (point-to-multipoint transmission) employing an Multimedia Broadcast/Multicast Service (MBMS).  We investigate two metrics to gauge the performance of MBMS/MBSFN transmissions in comparison with standard unicast transmissions for vehicular communications: latency of packet delivery and overhead caused by vehicular traffic, i.e., network utilization. Additionally, we present technique of prediction of system behaviour and explore the influence of transmission bandwidth and transmission rate on mentioned metrics. 
\end{abstract}
\acresetall
\begin{IEEEkeywords}
Multicasting, MBSFN, Vehicular Communications, V2X, Latency.
\end{IEEEkeywords}
\vspace*{-15pt}
\section{Introduction}
\label{sec:Introduction}

Multimedia Broadcast/Multicast Service (MBMS) was introduced by the 3rd Generation Partnership Project (3GPP) as means to broadcast and multicast information to 3G and 4G mobile users, with mobile TV being the main service offered \cite{3gpp.25.346,3gpp.22.146}. In the context of LTE systems, MBMS was evolved into e-MBMS increasing the performance of the air interface with a new transmission scheme called Multicast-Broadcast Single-Frequency Network (MBSFN). In MBSFN operation, MBMS data are transmitted simultaneously from multiple strictly time and frequency synchronized cells. A group of such cells transmitting these data establishes the so-called MBSFN area \cite{3gpp.36.300}. The increase in performance of the air interface is obtained in MBSFN due to great enhancement in the Signal to Interference and Noise Ratio (SINR) which is especially beneficial for the users at cell edge \cite{toskala2009lte}.\\
Substantial developments have taken place over the past few years in the area of vehicular communication systems. After the deployment of various vehicular technologies, such as toll collection or active road signs, vehicular communication (VC) systems have emerged. These systems include network nodes, that is, vehicles and road-side infrastructure units (RSUs) equipped with onboard sensory, processing, and wireless communication modules \cite{papadimitratos2008secure}. Vehicle-to-infrastructure (V2I) and vehicle-to-vehicle (V2V) communication allows a range of applications to increase transportation safety and efficiency, as well as video streaming \cite{7018153}. Especially road safety applications play a very important role in vehicular communications. Road safety applications rely on short-message broadcasting in a vehicle's neighbourhood to inform other vehicles in order to reduce accidents on the road. As a new traffic model, these applications exhibit some unique features in terms of generation patterns and delivery requirements. Particularly, delivery requirements of road safety applications are of high importance, since any signal delay increases the danger of accidents. Additionally, the possibility to support ordinary unicast users is highly preferable. In order to sustain such requirements, MBSFN can be considered as a potential way to handle vehicular applications \cite{valerio2008umts}. \\
The remainder of this paper is organized as follows. In Section~\ref{sec:System Model} we define MBMS/MBSFN transmission and explain our traffic model. Our performance metrics, i.e., the latency definitions as well as the network utilization are discussed in Section~\ref{metrics1}. Next in Section~\ref{sec:CQI}, methods for adaptation of Channel Quality Indicator (CQI) are motivated and explained. Final simulation results and discussions are described in Section~\ref{sec:Simulations}. Concluding remarks are  provided in Section~\ref{sec:Conclusion}.
\begin{figure}
\centering
\begin{center}
\captionsetup{justification=centering}
\includegraphics[scale = 0.45]{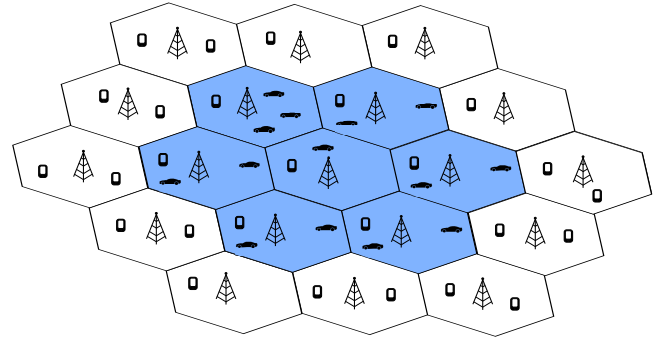} 
\caption{Illustration of a cellular network serving vehicles via LTE's MBMS/MBSFN feature as well as ordinary static users over unicast transmission.}
\vspace{-12pt}
\label{fig:sim}
\end{center}
\end{figure} 
\vspace*{-5pt}
\section{System Model}
\label{sec:System Model}
\begin{figure}
\centering
\begin{center}
\captionsetup{justification=centering}
\includegraphics[scale = 1.1]{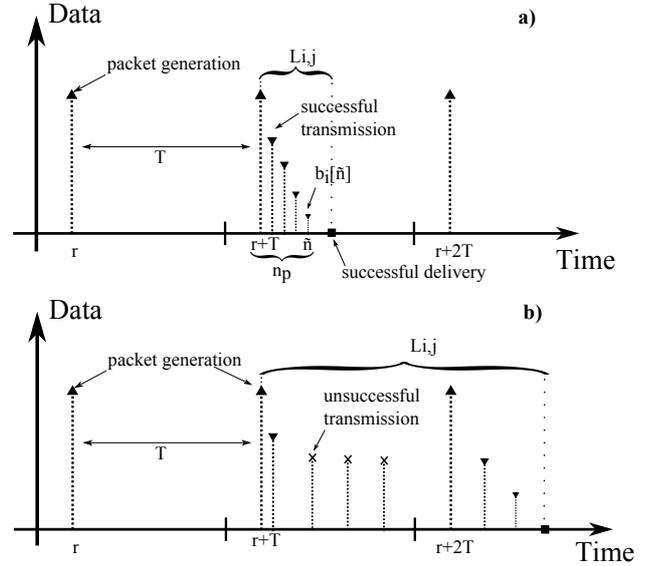} 
\caption{Explanation of latency calculation.}
\vskip-20pt
\label{fig:latency}
\end{center}
\end{figure}

We consider Single-Input Single-Output (SISO) transmission in the downlink of a cellular network. The transmitter employs OFDM modulation to  convert  the  frequency  selective  channel  into  a  set  of  non-interfering frequency-flat subcarriers indexed by $n$. The input-output relationship of user $i$ at subcarrier $n$ in case of MBSFN transmission is 
\begin{equation}
y_i[n] = \sum_{j\in\mathrm{MBSFN}}h_{i,j}[n]\cdot x_{\mathrm{MBSFN}}[n]+\sum_{l\not\in\mathrm{MBSFN}}h_{i,l}[n]\cdot x_l[n]+z_i[n]
\label{in-out-mult}
\end{equation}
where $j$ denotes the base station index in the MBSFN area,  $x_{\mathrm{MBSFN}}[n]$ denotes MBMS data, which is the same for all multicast users, $z_i[n]$ is Additive White Gaussian Noise (AWGN) and $h_{i,j}[n]$ is complex channel coefficient which can be expressed as  $h_{i,j}[n] = \gamma_j\cdot \widetilde{h}_{i,j}[n]$, where $\gamma_j$ denotes macroscopic pathloss and shadow fading and $\widetilde{h}_{i,j}[n]$ represents microscopic fading.\\
For standard unicast transmissions the input-output relationship for user $i$ can be expressed as 
\begin{equation}
y_i[n] = h_{i,j}[n]\cdot x_{\mathrm{MBSFN}}[n] + \sum_{l\neq j} h_{i,l}[n]\cdot x_l[n] + z_i[n]
\label{in-out-uni}
\end{equation}
where $x_{\mathrm{MBSFN}}[n]$ denotes the MBMS data to be transmitted.
Based on (1) we can expressed the SINR of MBMS user $i$ in case of multicast transmission as
\begin{equation}
\mathrm{SINR}_{i,\rm multicast} = \frac{|\sum_{j\in\mathrm{MBSFN}} h_{i,j}|^2}{\sigma_z^2 + \sum_{l\not\in\mathrm{MBSFN}}| h_{i,l}|^2}~.
\label{SINR-mult}
\end{equation}
Similarly, based on (2) the SINR of MBMS user $i$ in case of unicast transmission can be expressed as
\begin{equation}
\mathrm{SINR}_{i,\rm unicast} = \frac{|h_{i,j}|^2}{\sigma_z^2 + \sum_{l\neq j}| h_{i,l}|^2}~.
\label{SINR-uni}
\end{equation}
In our work we assume delay- and error-free uplink transmission from vehicles to base stations and mainly focus on the downlink domain. According to Figure \ref{fig:latency} we assume that each car user generates MBMS data of size $p_S$ bits at random starting time $r$ and then produces packets every $T$ ms. Such packet generation can be observed, for example, in road-safety applications, when  cooperative awareness messages (CAM) \cite{etsi2010302} are generated. These data should be successfully distributed to all other vehicles in MBSFN area via multicasting or unicasting.
The buffer size of car user $i$ at time $\widetilde n$ can be calculated as
\begin{equation}
b_i[\widetilde n]  = p_s-\sum_{m=1}^{n_p}p_t[\widetilde{n}-m]
\label{buffer_size}
\end{equation}
where $p_t[\widetilde{n}-m]$ is successfully transmitted packet of size $p_t$ bits at time $[\widetilde{n}-m]$ and $n_p$ is specified as
\begin{equation}
 n_p =
  \begin{cases}
   n-\lfloor\frac{n}{T}\rfloor T-r &, r<\widetilde{n},\quad \widetilde{n} = n-\lfloor\frac{n}{T}\rfloor T \\
   n-(\lfloor\frac{n}{T}\rfloor-1)T-r&, r>\widetilde{n},\quad \widetilde{n} = n-\lfloor\frac{n}{T}\rfloor T\\
  \end{cases}
\end{equation}\\
which denotes the time difference between packet generation and time instance $\widetilde{n}$.
We consider latency as time interval between data generation and successful delivery to all appropriate users within MBSFN area. 
The latency value can be calculated for both cases of erroneous and error free transmission. According to LTE standard the Hybrid automatic repeat request (HARQ) is not specified in MBMS transmissions. It means that in case of the unsuccessful transmission of a packet, the packet will not be retransmitted but instead we accumulate the latency until we successfully receive the next packet from the same user.  Additionally, if during waiting time new packets were generated, the old packets replace them. 
In Figure \ref{fig:latency} two cases of latency calculation are shown in more details. In the upper part of Figure \ref{fig:latency}, the latency evaluation in case of success transmission is described, while in the lower part the procedure of latency accumulation in case of unsuccessful transmission is explained. 
Assuming total number of MBMS users equal  $N_{\text{m\_ue}}$ and each of them generate in total $N_{\text{packets}}$ packets, we stack corresponding  latency values into a large matrix $\boldsymbol{L}$ of size $ N_{\text{packets}}\times N_{\text{m\_ue}}$, with elements  
\begin{equation}
L_{s,i} =
\begin{cases}
 t_i  &, \text{error free transmission} \\
 t_i+T\cdot k &, \text{errorneous transmission} \\
 \end{cases}
\end{equation}
where $t_i$ is the time when $b_i[t_i] = 0 $, i.e. complete transmission of packet $s$ and $k$ is the number of required retransmissions.
\section{Performance Metrics}
\label{metrics1}
\subsection{Latency Evaluation}
During our investigation we came to the conclusion that the latency evaluation should be performed highly accurately and transparently since interpretation mistake of one of the most important parameters in vehicular communication leads to improper decisions in network specification. Thereby we define three different ways of latency performance indicators:\\ \textbf{Combined latency CDF}: We transform matrix $\boldsymbol{L}$ into a vector $\mathbf{\widehat{L}}$ of size $N_{\text{packets}}\cdot N_{\text{m\_ue}}\times 1 $ and calculate the empirical cumulative distribution function (ECDF)
\begin{equation}
\mathrm{CDF}_{\rm combined} = \mathrm{ECDF}\big(\mathbf{\widehat{L}}\big)~.
\end{equation}
It should be noticed that the main contribution to this latency evaluation is added by the users that have high SINR and, as a consequence, represent significant amount of low latency receptions.\\
\textbf{CDF of mean latency}: We determine the mean latency for each user position (average over all latency values $s$ obtained at a given user $i$) and calculate the CDF of these mean latencies.
\begin{equation}
\mathrm{CDF}_{\rm mean} = \mathrm{ECDF}\big(\mathbf{\widetilde{L}}\big)
\end{equation}
where ${\widetilde{L}}_i = \frac{1}{N_{\text{packets}}}\sum_{s = 1}^{N_{\text{packets}}}L_{s,i}$. Notice that this method does not represent the worst latency, which is, however, a critical indicator especially for safety-relevant applications (road-safety transmission).\\
\textbf{Latency of individual users}: We determine the latency ECDF of each user position individually, i.e., we obtain $ N_{\text{m\_ue}}$ CDFs corresponding to different car users within the network.
\subsection{Network Utilization}
Network utilization is considered as another important performance metric which gives us better understanding of the price to be paid in terms of throughput of ordinary unicast users for supporting MBSFN transmission.
In our investigation we evaluated the network utilization as a percentage of resources to be used for sustaining the MBMS traffic. It can be calculated as
\begin{equation}
\mathrm{Util}= \frac{p_s\cdot  N_{\text{\rm m\_ue}}}{N_{\text{RB}}\cdot N_{\text{RE}} \cdot \mathrm{Efficiency}_{\mathrm{CQI[n]}} }\cdot 100\%
\label{util}
\end{equation}
where $N_{\text{RB}}$ is a number of resource blocks, $N_{\text{RE}}$ denotes number of recourse elements per resource block and $\mathrm{Efficiency}_{\mathrm{CQI[n]}}$ is the efficiency of the CQI (in bits per resource element) chosen for transmission of MBMS data.  With (10) we can calculate the appropriate number of subframes to be reserved for MBMS data transmission thus satisfying the trade off between MBSFN subframes and subframes used for supporting of ordinary users. Additionally it helps us not to go into network congestions and avoid reservation of excessive number of subframes for MBSFN transmission which is beneficial for the throughput of ordinary unicast users. Network congestion in our meaning indicate the situation when the old message from specific user in not delivered to all MBSFN users while the new message is already generated. This situation has an avalanche effect which leads to significant degradation of the system in terms of latency.
\label{sec:Metrics}
\section{CQI Adaptation}
\label{sec:CQI}
In Section \ref{sec:Simulations} we present system performance with and without rate adaptation. Irrespective whether rate adaptation was applied, we reserve the same number of subframes for MBMS data, calculated with  (10). However, for rate adaptation the real amount of used subframes could be reduced and unused MBMS subframes can instantaneously be reassigned for ordinary traffic, which may not be feasible in practice. The CQIs of all users in the MBSFN area are calculated according to proposed in \cite{schwarz2011throughput} technique and stored in the vector $\boldsymbol{\mathrm{CQI}}$. From the vector $\boldsymbol{\mathrm{CQI}}$ we choose the smallest CQI index for transmission, in order to support all users. During our research we found that using the smallest CQI index for transmission can cause traffic congestions (since the number of reserved subframes for MBSFN transmission is too small for supporting communication with such a low efficiency) and we should specify some lower bound ($\mathrm{CQI_{\rm bound}}$) to assure that the generated MBMS traffic can be sustained by the network. Therefore the CQI index at time $n$ to be used for transmission can be calculated as
\begin{equation}
\mathrm{CQI[n] = \max(\min_i(\boldsymbol{\mathrm{CQI}}[i]), CQI_{\rm bound})}~.
\end{equation}
\section{Simulations}
\label{sec:Simulations}
\begin{figure}
\begin{subfigure}{\columnwidth}
\centering
\captionsetup{justification=centering}
\includegraphics[scale=0.25]{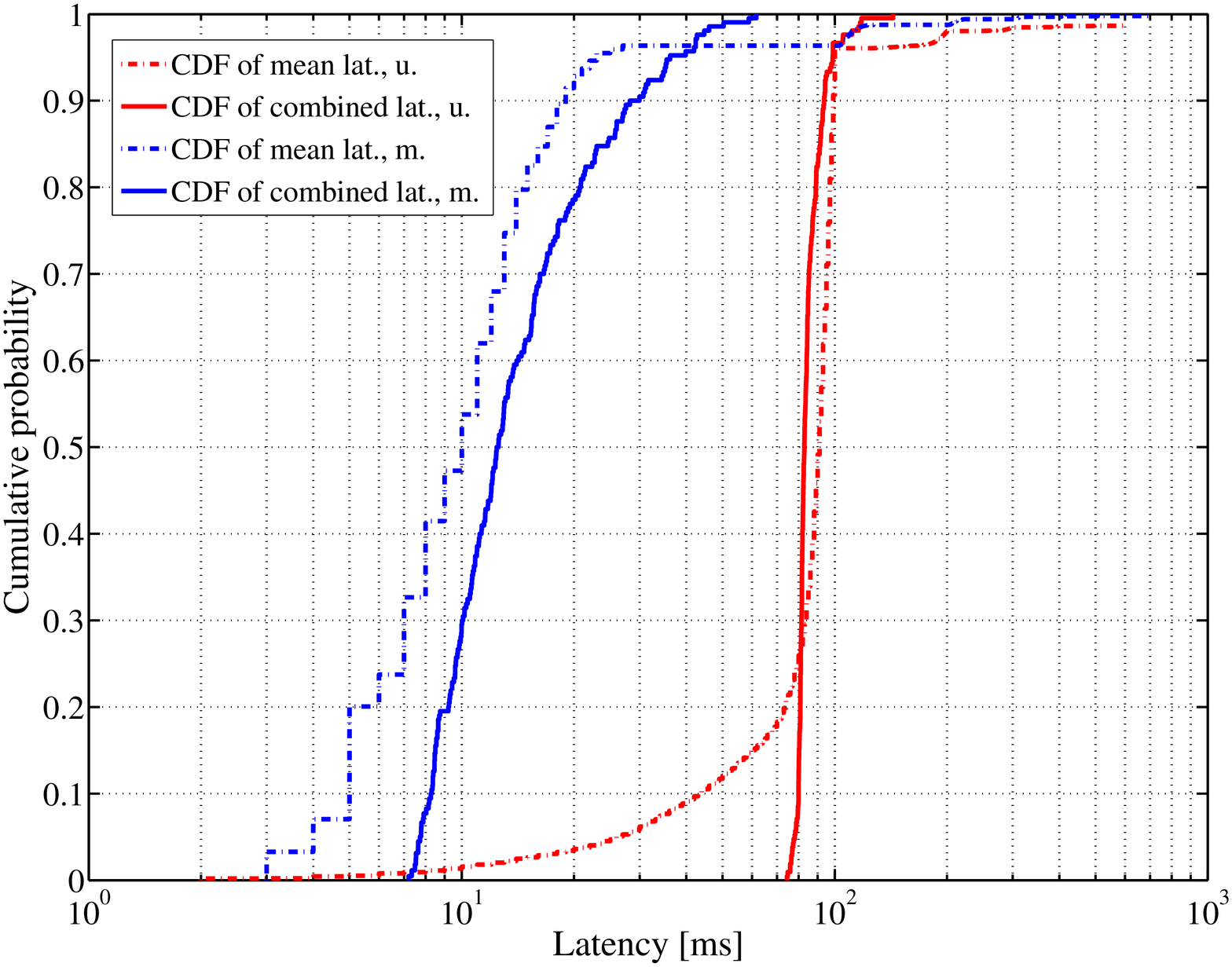}
\caption{Comparison of CDF of different latencies for unicasting and multicasting of CAMs at 5MHz bandwidth and transmission with CQI3.}
\label{fig:lat_comp1}
\end{subfigure}
\begin{subfigure}{\columnwidth}
\centering
\captionsetup{justification=centering}
\includegraphics[scale=0.25]{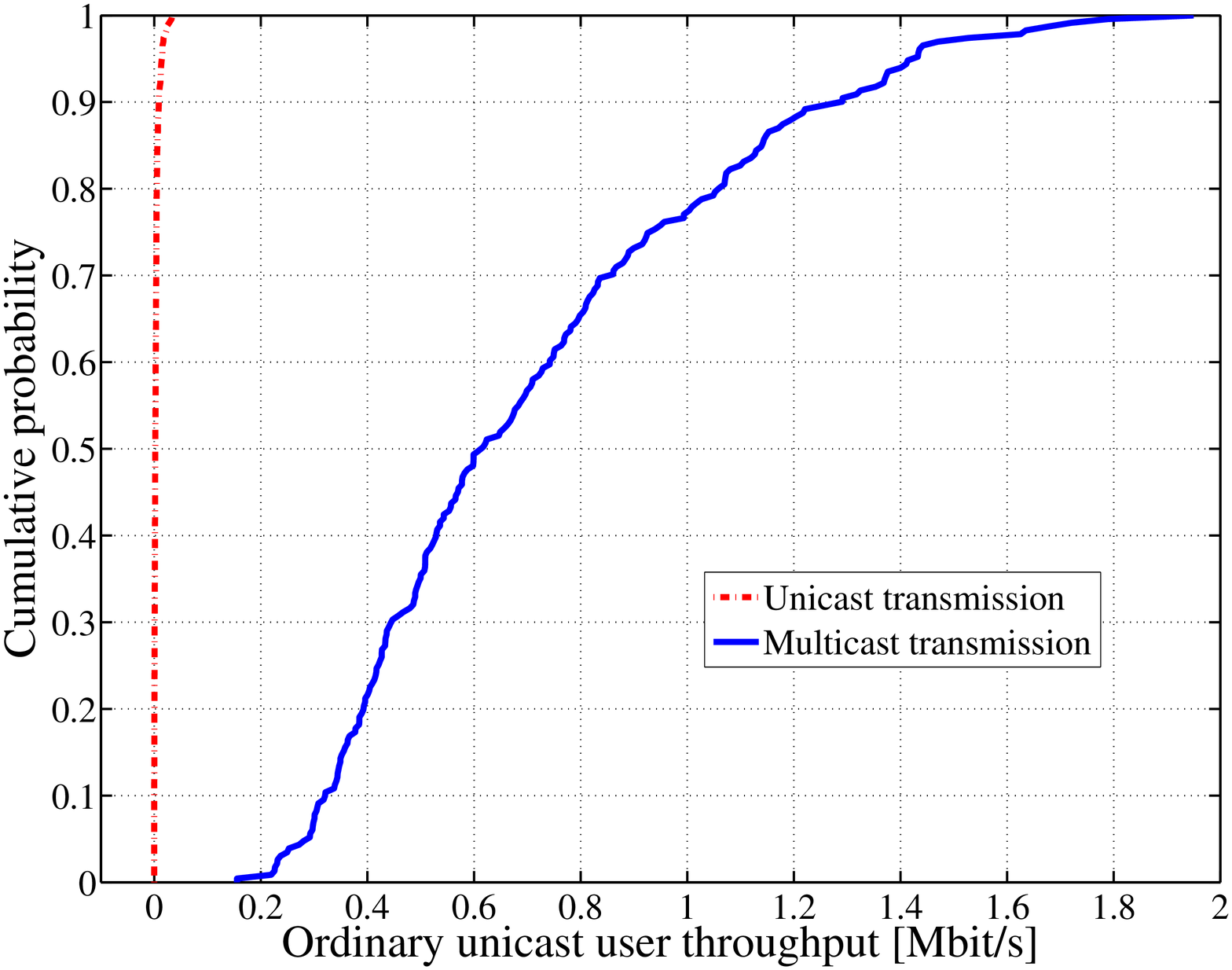}
\caption{Comparison of CDF of ordinary user throughput at 5MHz bandwidth.}
\label{fig:thr_compp}
\end{subfigure}
\caption{Comparison of the system performance in terms of latency and ordinary unicast user throughput between multicasting and unicasting of CAMs with CQI3.}
\vspace{-10pt}
\end{figure}

Our simulations were carried out with the Vienna LTE System Level Simulator \cite{schwarz2013pushing}, where we consider an MBSFN area with several high mobility users ("cars") and ordinary unicast users as illustrated in Figure 1. The car users generate CAMs of size $p_s = 300$ bytes randomly in interval of $T = 100$ ms. These messages should be distributed among cars within the MBSFN area. MBMS data should be transmitted in reserved subframes and the standard unicast full buffers users are served with the remaining resources.\\
After this explanation the discussion of the size of MBSFN area arises. It is clear that with including more and more base stations in the MBSFN area we decrease the number of potential interference sources which leads to improvement of SINR. Nevertheless the large size of the MBSFN area can cause high delay echoes which can further reduce the performance of the system introducing inter-symbol interference (ISI). One way to consider the effects of ISI is proposed in \cite{erich+stef}, where the authors  introduce an extended feedback algorithm, which accounts effects of ISI in high delay systems. It should be added that the amount of multicast users will increase with including more base stations and, as a consequence, the amount of generated data will also increase. This data "explosion" could cause undesirable congestions in the network and significant delays in transmission due to buffer overflows. Also with increasing the MBSFN area we increase the number of recipients for which the information could be irrelevant. All these circumstances impact the preferred size of the MBSFN area. \\
We decided to use in our simulations structure depicted in Figure 1. In this case the MBSFN area of seven base stations is surrounded by a ring of interference base stations. This allows us to simultaneously simulate both practical conditions and beneficial features of broadcasting information while avoiding network congestions.\\
\begin{table}[t]
\renewcommand{\arraystretch}{1.3}
\caption{Simulation Parameters}
\centering
\begin{tabular}{r l}
\hline
\textbf{Parameter}				&		\textbf{Value}	\\
\hline
Center frequency                &      2.14GHz            \\ 
System bandwidth				&		5, 20MHz	         \\
Channel                         &   ITU-T VehA \cite{Veh-A}             \\
Number of base stations in MBSFN&		7		          \\
Number of users per base station& 		6	\\
Number of car users per base station	&		3	\\
Speed of car users              & 100 km/h\\
Transmission                    &   unicast / multicast  \\
Packet size to be transmitted   & 300 bytes\\ 
Packet generation rate (T)    &  10Hz  \\
Transmission rate               & Rate adaptation /fixed rate for car users\\
                                & Rate adaptation for ordinary unicast users\\
Antenna configuration           & $1\times 1$\\
\hline 
\label{table:sim}
\end{tabular} 
\end{table} 
The main parameters in the simulations are presented in Table \ref{table:sim}. In Figure \ref{fig:lat_comp1} we compare the CDF of mean latencies for unicast and multicast transmissions of CAMs. We observe significant advantages of multicast transmissions comparing to the unicast case in terms of latency  which  is even more remarkable in terms of throughput of standard unicast users in Figure \ref{fig:thr_compp}. Such throughput reduction in case of unicasting is explained by resource consumption: transmission of CAMs via unicasting consumes 99.95\% of cell resources, while multicasting consumes 52\%. Hence, through this paper we investigate mainly the behaviour of multicast transmissions in MBSFN area.
\subsection{Bandwidth Scaling}
\begin{figure}
\begin{subfigure}{\columnwidth}
\captionsetup{justification=centering}
\includegraphics[scale=0.25]{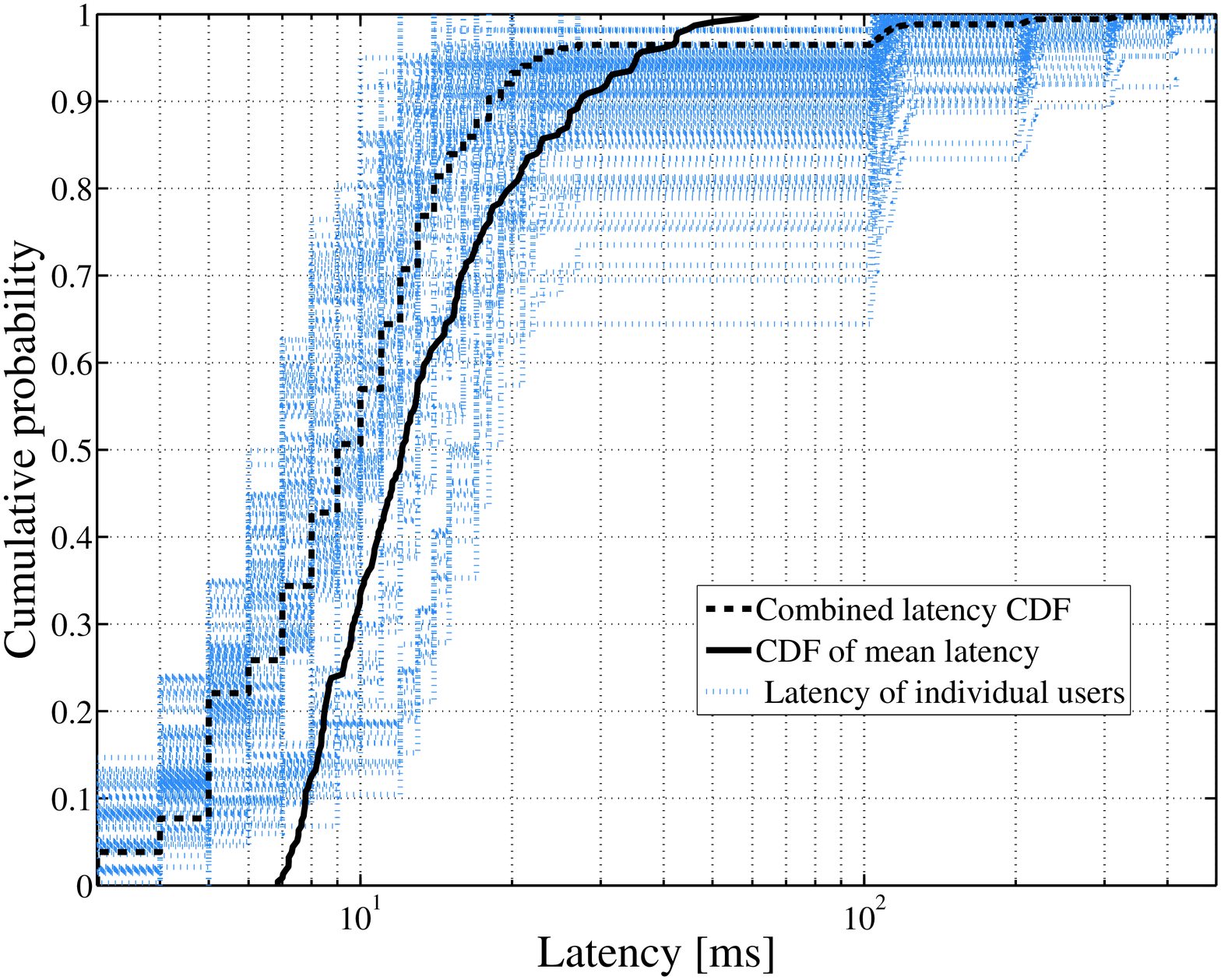}
\caption{Comparison of CDF of latency for multicast transmission at 5MHz bandwidth.}
\label{5MHzscaling}
\end{subfigure}
\begin{subfigure}{\columnwidth}
\centering
\captionsetup{justification=centering}
\includegraphics[scale=0.25]{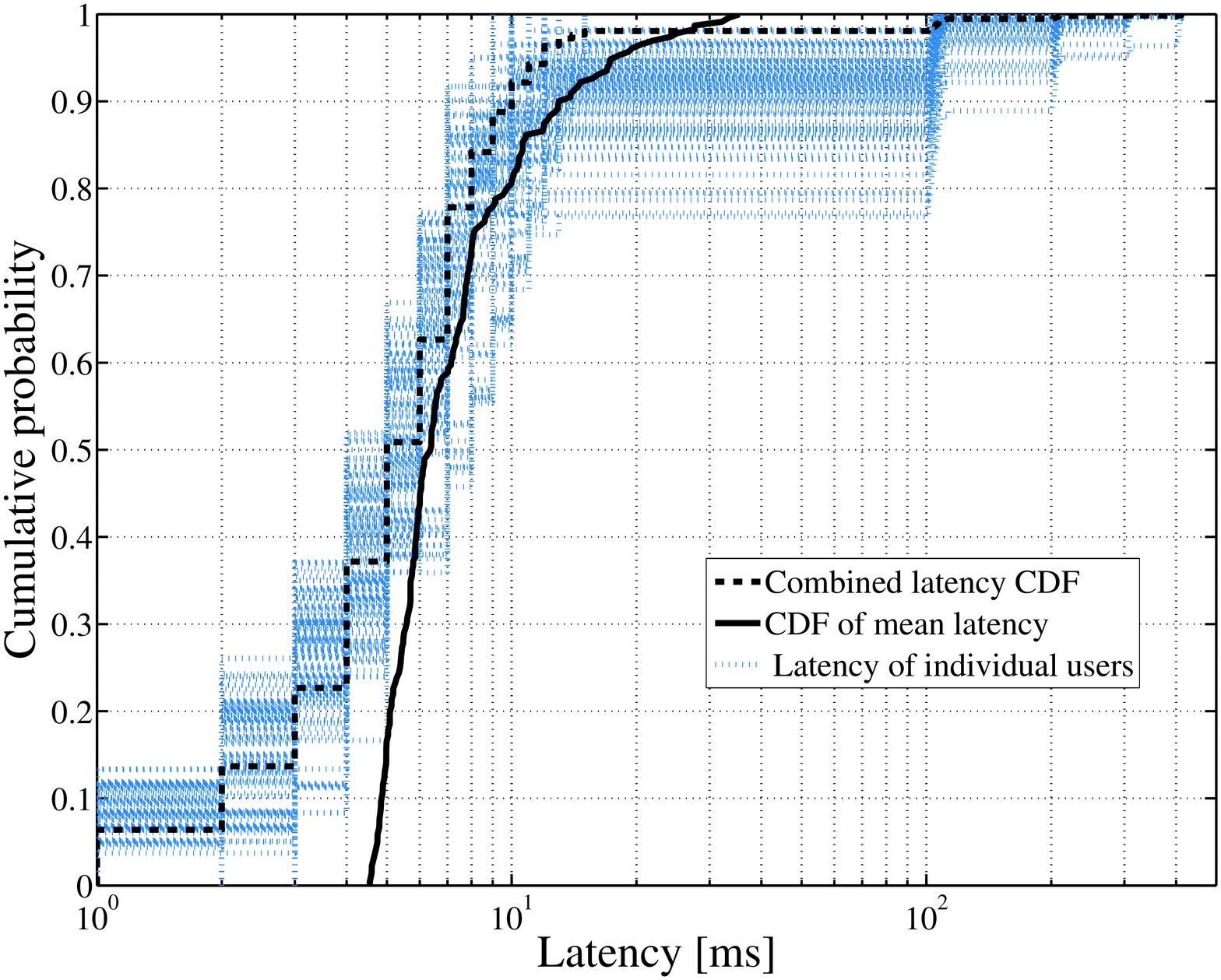}
\caption{Comparison of CDF of latency for multicast transmission at 20MHz bandwidth.}
\label{20MHzscaling}
\end{subfigure}
\begin{subfigure}{\columnwidth}
\centering
\captionsetup{justification=centering}
\includegraphics[scale=0.25]{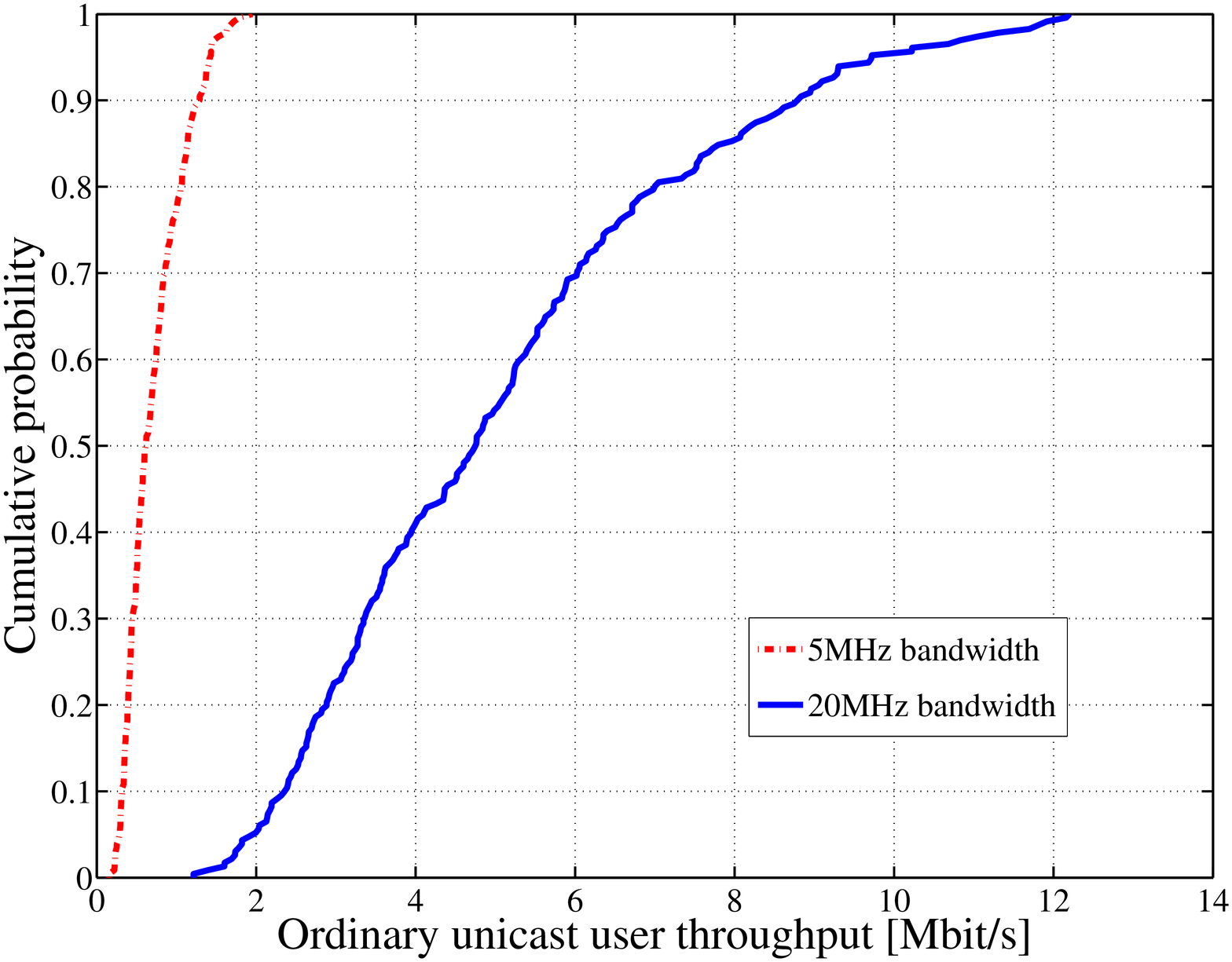}
\caption{Comparison of CDF of ordinary user throughput at 5MHz bandwidth and 20MHz bandwidth.}
\label{fig:thr_comp2}
\end{subfigure}
\caption{Impact of the transmission bandwidth on latency and ordinary unicast user throughput.}
\end{figure}
In Figure \ref{5MHzscaling} and \ref{20MHzscaling}  we compare the latency of multicast CAM transmission for both considered system bandwidths of 5 and 20\,MHz and transmission with CQI3 (efficiency is 0.377). With 5MHz bandwidth we have to reserve six subframes per radio-frame for MBMS transmission. However, if not all of them are required, we reassign unused MBMS subframes for unicast transmission. With 20MHz bandwidth, two subframes are sufficient. The throughput comparison of ordinary unicast users is shown in Figure \ref{fig:thr_comp2} and the corresponding mean values are provided in Table 2. Given the multicast network utilization values from Table 2, the expected throughput improvement is:
$$\frac{R_{20}}{R_5}= \frac{(1-0.157)\cdot 100}{(1-0.52)\cdot 25} = 7.03$$
where $R_{20}$ and $R_{5}$ are the number of resource blocks utilized for serving ordinary users at 20MHz and 5MHz bandwidth. According the Table 2, the observed improvement equals 7.08. Hence, the throughput values scale very well with the bandwidth, provided the network utilization is considered, which allows us to predict the impact of bandwidth and number of MBMS users on system behaviour and especially on the achievable mean throughput of ordinary unicast users.
\begin{table}[h]
\renewcommand{\arraystretch}{1.3}
\centering
\captionsetup{justification=centering}
\begin{tabular}{c| c c c}
\hline
Bandwidth			&		mean latency [TTI] & mean throughput [Mbit/s] & utilization[\%]	\\
\hline 
5MHz               &     14.7                &      0.72                  &          52 \\
20MHz          &             8                  &    5.1               &             15.7
 \vspace{-5pt}
\label{table:results2}
\end{tabular}
\caption{Summary of performance results with 5 and 20MHz bandwidth and transmission with CQI3.}
\vspace{-20pt} 
\end{table}
\subsection{Rate Adaptation for Multicast Users}
\begin{figure}[ht!]
\begin{subfigure}{\columnwidth}
\centering
\captionsetup{justification=centering}
\includegraphics[scale=0.25]{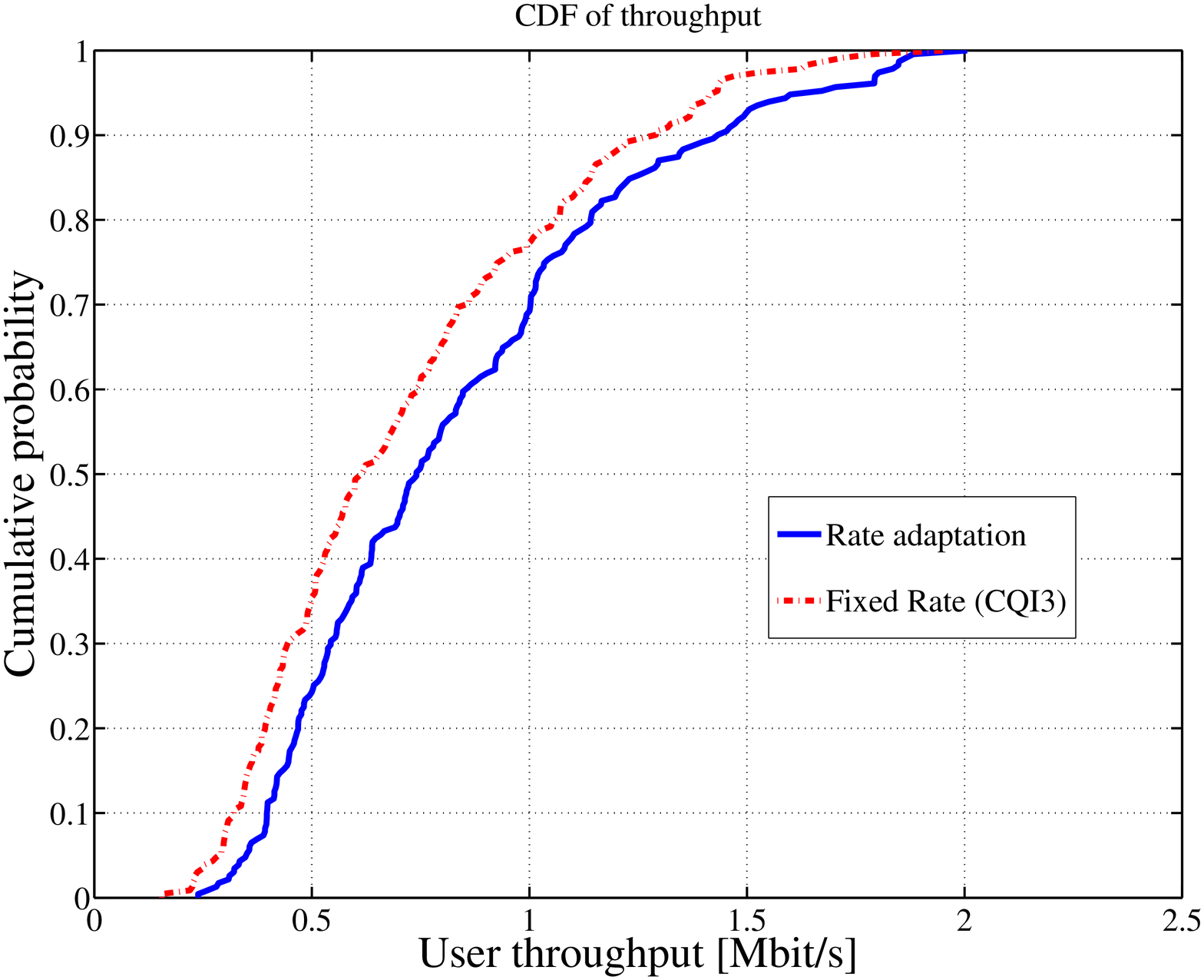}
\caption{Comparison of the throughput of ordinary unicast users with fixed rate and rate adaptive multicasting.}
\label{fig:lat_comp}
\end{subfigure}
\begin{subfigure}{\columnwidth}
\captionsetup{justification=centering}
\includegraphics[scale=0.25]{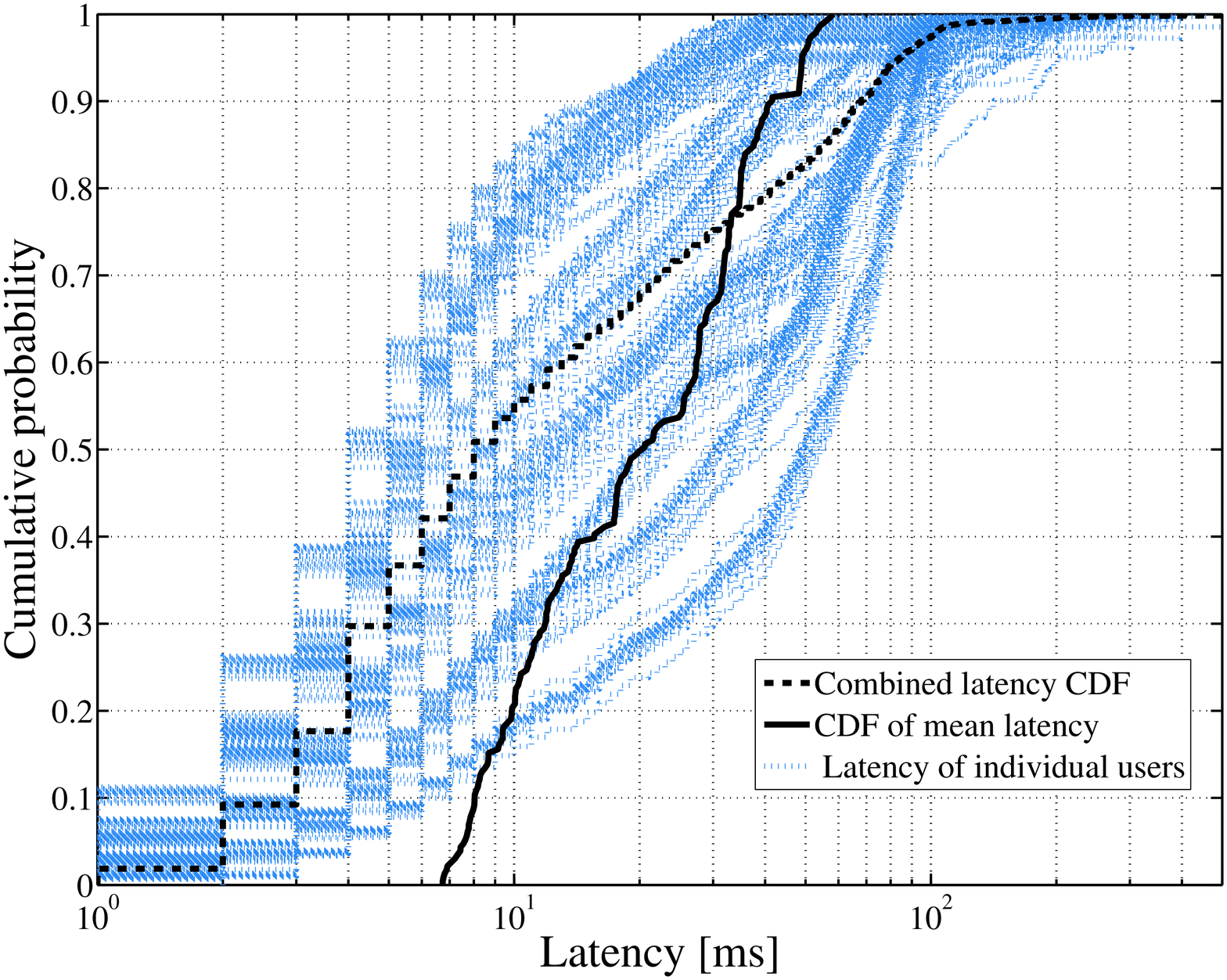}
\caption{Latency with 5MHz and no CQI lower bound.}
\label{fig:now_bound}
\end{subfigure}
\begin{subfigure}{\columnwidth}
\centering
\captionsetup{justification=centering}
\includegraphics[scale=0.25]{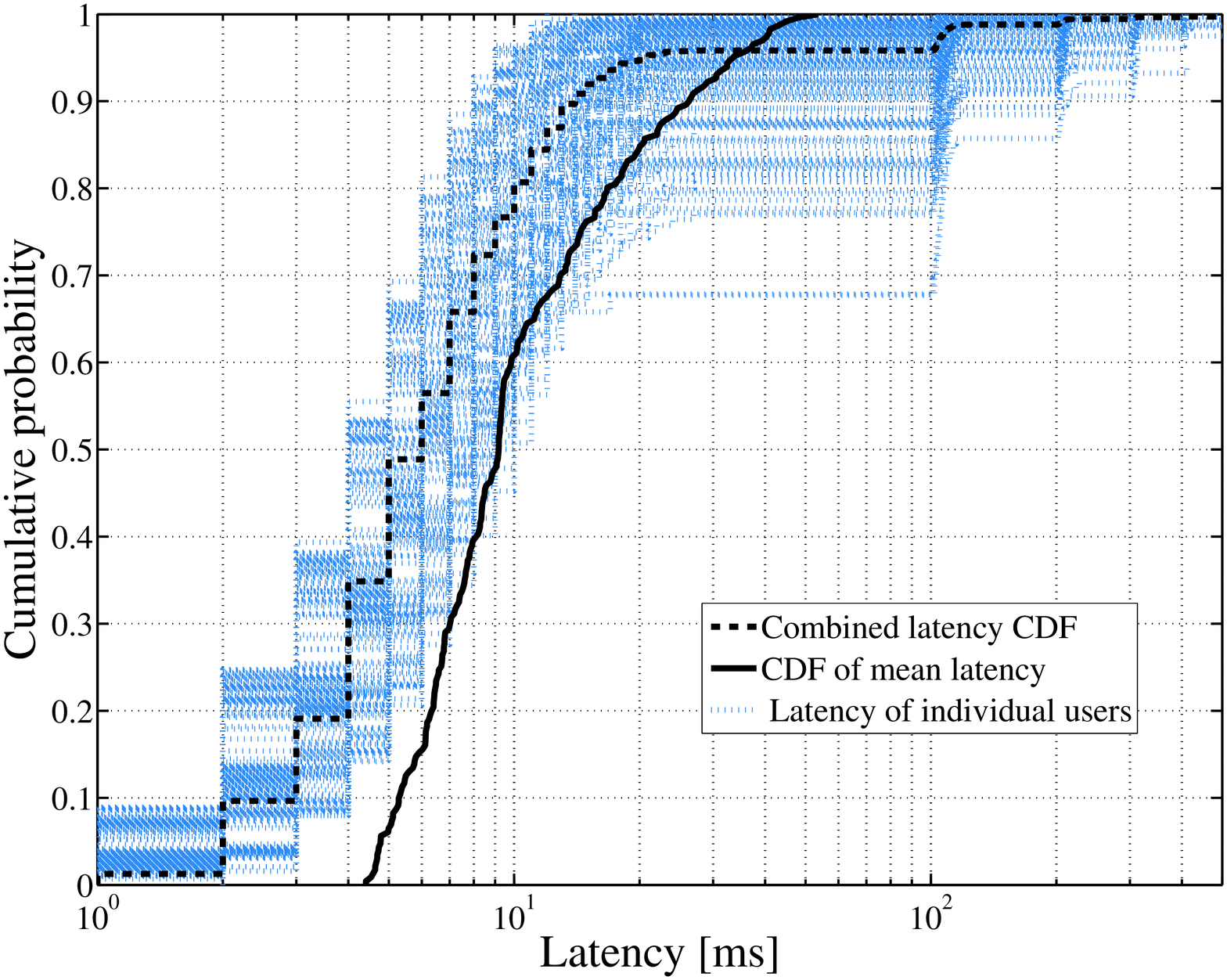}
\caption{Latency with 5MHz bandwidth and CQI lower bound of three.}
\label{fig:thr_comp}
\end{subfigure}
\label{fig:rate_adapt}
\caption{Impact of rate adaptation on latency and ordinary unicast user throughput.}
\vspace{-15pt}
\end{figure}

We now consider transmissions with rate adaptation for multicast users. At first, we investigate how the performance changes if we do not apply a lower bound for the CQI, i.e., if we simply take the minimum CQI of all users even if we cannot sustain the traffic in this way. This is shown in Figure \ref{fig:now_bound}. Then we perform rate adaptation with the lower bound of CQI 3. Corresponding results are illustrated in Figure 5 and summarized in Table 3. According to Figure 5b and 5c we can see that in the case of rate adaptation with lower bound the deviation of CDF of individual latencies is much smaller. In such systems the performance is determined by the worst users and on Figure 5c we can see that in average employing lower bound leads to better results.  We observe an improvement in the mean latency by a few TTIs with rate adaptation, which can be explained by the fact that we now require less RBs for transmission in case the CQI of all users is high (exploiting channel diversity). This can also be seen in the average network utilization for multicast transmission, which reduces by almost 10\%. Correspondingly, the mean throughput of ordinary users improves from 0.72 Mbit/s to 0.83 Mbit/s. 
\begin{table}
\renewcommand{\arraystretch}{1.3}
\centering
\captionsetup{justification=centering}
\begin{tabular}{c| c c c}
\hline
transmission rate			&		mean latency [TTI] & mean throughput [Mbit/s] & utilization[\%]	\\
\hline 
CQI3             &     14.7                &      0.72                  &          52 \\
adaptive         &             11.9        &    0.83                    &         43
\label{table:results3}
\end{tabular}
\caption{Performance results for fixed CQI and adaptive CQI transmission with 5MHz bandwidth.} 
\vspace{-15pt}
\end{table}
\vspace{-5pt}
\section{Conclusion}
\label{sec:Conclusion}
In this paper, we investigated the performance of LTE MBSFN networks in terms of cell resource utilization, latency of packet delivery and throughput of ordinary unicast users. We observed significant advantage of multicast over  unicast transmission for supporting vehicular communication. Additionally, we introduced several concepts of latency evaluation and proposed a technique for system behaviour prediction in case of different transmission bandwidth. It should be mentioned that the full picture of packet delivery latency is only provided if we do not apply any kind of latency aggregation over users and/or time. With rate adaptation technique we can achieve smaller packet delivery time and higher throughput, however we should include a minimum decision boundary to sustain the MBMS traffic. \vspace{-8pt}
\bibliography{my}
\bibliographystyle{IEEEtran}
\end{document}